# Bankruptcy as an exit mechanism for systems with a variable number of components

Corrado Di Guilmi,[a,*] Edoardo Gaffeo,[b] Mauro Gallegati,[a]

[a] *Department of Economics and WEHIA2, Università Politecnica delle Marche,*
*Piaz.le Martelli 8, I-60121 Ancona, Italy*

[b] *Department of Economics and CEEL, University of Trento,*
*Via Inama 5, I-38100 Trento, Italy*

**Abstract**

Dynamical systems with components whose sizes evolve according to multiplicative stochastic rules have been recently combined with entry and exit processes. We show that the assumptions usually made in modeling exits are at odds with the available evidence. We discuss a recently proposed macroeconomic model with random multiplicative shocks and a mechanism for exit based on bankruptcy, which displays several observed stylized facts for firms' dynamics, like power law distributions for firms' sizes and Laplace distributions for firms' growth rates.

*Keywords*: Power law, bankruptcy, Econophysics

## 1. Introduction

One of the much studied topics inside the Econophysics [1] community is the emergence of scaling or power law relationships. Indeed, the power law distribution has proved to be a useful concept in the study of financial markets [2], industrial dynamics [3], macroeconomic fluctuations [4], urban economics [5] and social network formation [6]. Several generative models have been proposed so far, ranging from self-organized criticality [7] to random *killing* processes [8]. Among them, economists have been particularly attracted by random multiplicative processes, basically because of their acquaintance with the pioneering work of R. Gibrat [9], who proposed the so-called *Law of Proportional Effect* (i.e. growth rates uncorrelated and independent of size) to modeling the dynamics of firms' size back in the 1930s.

Since it is well known that the stochastic processes based on the Gibrat's Law generate a log-normal distribution, much effort has been recently pursued to reconcile theory with the evidence of a power law distribution for firms size [10, 11]. In particular, recent research [12, 13] has shown that power laws naturally emerge from systems where a random multiplicative process is combined with the entry end exit of systems' components. While simulations show that results are robust to different specifications for entry and exit mechanisms, scarce attention has been paid so far to understand how realistic such processes are.

---

• Corresponding author.
  E-mail address: diguilmi@dea.unian.it (C. Di Guilmi).

This paper aims to partially address this issue, showing that a particular feature holds for firms' exit: the debt of bankrupted firms (*bad debt*, in the common jargon) – i.e. one of the causes for firms going out of business – scales as a power law. The evidence we report for bankrupted firms in Italy, Spain and France and a sample of bankrupted European bond issuers is strikingly similar to that found for Japan [14, 15]. We further discuss a simple agent-based macroeconomic model based on financial fragility which, along with a power law distribution for firms size (and many other stylized facts), is capable to replicate this fact.

The remainder of this paper is organized as follows. Section 2 reports our findings of scaling and universality in the debt distribution of firms exited due to bankruptcy. Section 3 describes our model. Section 4 presents some conclusive remarks.

## 2. Evidence on the *bad debt* distribution

A firm can go out of business as an independent unit for three reasons: *i*) voluntary exit, due for example to the retirement of the founder entrepreneur; *ii*) exit due to the merger with another firm or the acquisition by another firm; *iii*) failure due to bankruptcy, here defined as the inability of a firm to pay its financial obligations as they mature.

In spite of the huge literature dealing with firms' exit available in economics, models of firm dynamics adding an exit process to multiplicative growth use rather unplausible exit mechanisms, that is they assume that firms exit as they reach a minimum size [12] or that the number of exits is time-invariantly proportional to the total population of firms [13]. Note that none of the reasons for exit listed above are by themselves fully compatible with neither the minimum size nor with the proportionality assumptions. In particular, exits due to bankruptcy – our topic of interest in this paper – occur because of a fatal deterioration of firms' financial conditions and large firms are far from immune, as the recent examples of Enron and MciWorldcom suggest. In fact, the available evidence [16] is clear-cut in suggesting that insolvences occur at all scales, and that the proportion of failures varies sensibly over time. Given the instrumental role of the exit process in replicating stylized facts on the size distribution of firms, this lack of consistency between modeling strategies and the lessons received from industrial economics might be seriously misleading.

With that in mind, we analyze the available evidence for bankruptcies in a sample of European countries, namely Italy, Spain and France. Data are retrieved from the commercially available dataset AMADEUS, which lists balance sheet data for about 6 millions European firms from 1992 through 2001. Information is also available on entry and exit of firms over time, as well as on the reason for exit.

First, we find that financial ratios are invariably a good predictor of firms failure, and therefore exit. In particular, the equity ratio, defined as the ratio between net worth (current assets minus current liabilities) and total assets, deteriorates almost monotonically as the date of bankruptcy approaches (Fig. 1(a)). The distribution of exits by age turns out to be in all cases exponential, signaling that the probability to fail is independent of time (Fig. 1(b)). Given that firms generally enter at a small scale, and that they grow over time through investments, this suggests that big and small firms should have a rather equal probability to go out of business.

From the viewpoint of debt, it is interesting to note that the right tail of the size distribution of the amount of debt of bankrupted firms $b$ scales down as a power law for all sampled countries (Fig. 3(a)),

$$Q(>b) \sim b^{-\alpha}. \tag{1}$$

In particular, the scaling exponent for the 60% right tails are $\alpha = -1.09$ for Italy, $\alpha = -0.87$ for France and $\alpha = -0.67$ for Spain. Furthermore, a quantitatively similar scaling exponent results also for data on defaulted debt collected from a different dataset, that is for the European corporate long term debt defaults occurred from January 1985 through May 2002, as reported by Moody's (Fig. 3(b)). The bad debt of insolvent bond issuers is distributed as a power law with $\alpha = -0.92$.

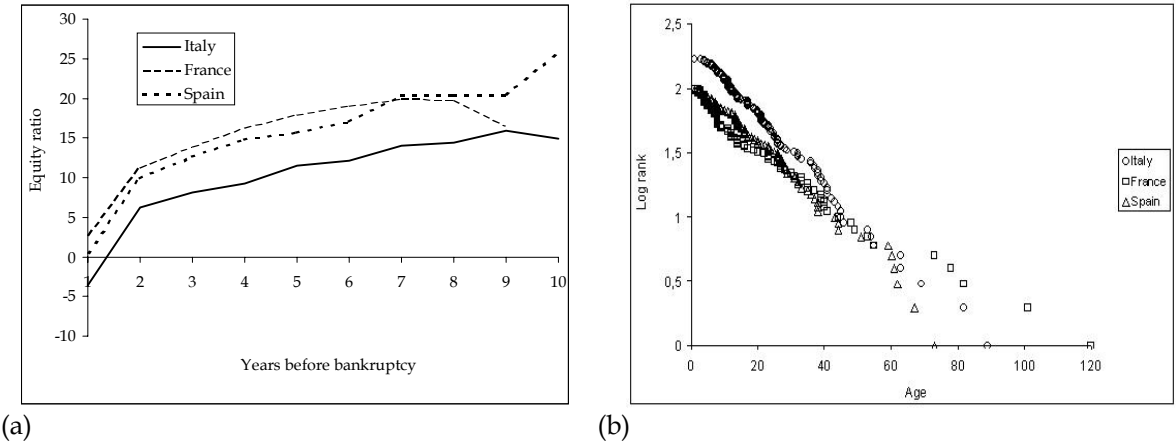

(a)                                                         (b)

Fig. 1. (a) Profile analysis of the average equity base of bankrupted firms in each of the year before bankruptcy. Data points are referred to mean values for 676 Italian, 1786 French and 750 Spanish firms went bankrupted during the 1992-2001 period. The data source is AMADEUS. (b) Semi-log plot of the distribution of bankrupted firms by age. We can see that for all countries in our sample the exponential distribution returns a good fit to the data.

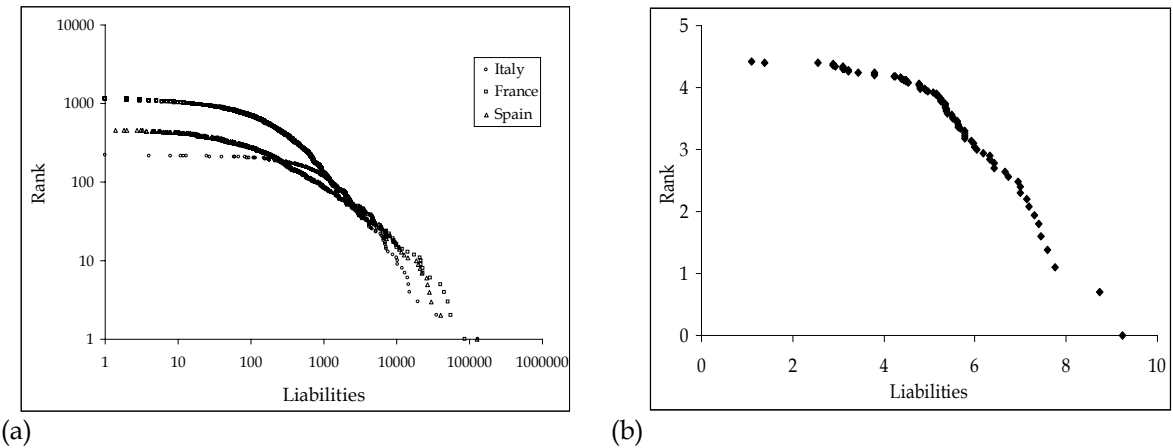

(a)                                                         (b)

Fig. 2. (a) Zipf plot for the debt of bankrupted firms in Italy (217 firms), France (1166), and Spain (455) during 1999. The data source is AMADEUS. Note that the right tails scale down as a power law. The scaling exponent for the 60% right tails are 1.09 ± 0.03, 0.87 ± 0.01 and 0.67 ± 0.01, respectively. (b) Zipf plot for European corporate long term debt defaults, from January 1985 through May 2002. The data source is Moody's. Also in this case the right tail scales down as a power law. The scaling exponent for the 60% right tail is 0.92 ± 0.03.

Our findings are strikingly close to the ones reported in [14, 15] for Japanese bankrupted firms, with the bad debt for large failed firms (i.e. the right tail of the distribution) being estimated to scale down with an exponent α comprised between 0.91 and 1. This result strongly suggests that universality, as defined in statistical physics, seems to hold for the bad debt distribution. Any reasonable model of industrial dynamics should take this evidence into account.

## 3. The model

A model [17] incorporating the assumptions of random multiplicative growth and exit due to bankruptcy has been recently proposed. The model consists of only two markets: goods and credit. The financial robustness of a firm is proxied by its equity ratio, that is the ratio of its equity base or net worth ($A$) to the capital stock ($K$), $a = A/K$. Since firms sell their output at an uncertain price they may fail. In particular, bankruptcy, and therefore exit, occurs when the net worth becomes negative, that is when the individual price falls below a critical threshold. The probability of bankruptcy turns out to be an increasing function of the capital stock, and a decreasing function of the equity base inherited from the past.

The dynamics is driven by the evolution of the capital stock, which in turn is determined by investment funded by bank credit. Each firm obtains a portion of total credit equal to its relative size: i.e. the ratio of the individual capital stock to the aggregate capital stock. Hence, rich (poor), i.e. highly (poorly) capitalized, firms obtain more (less) credit. The banking sector's equity base, which is the only determinant of credit supply, increases with banks' profits that are affected, among other things, by firms' bankruptcy. In fact, when a firm goes bankrupt the banking sector faces an insolvency, i.e. bad debt. This the root of a domino effect: the bankruptcy of a firm today is the source of the potential bankruptcies of other firms tomorrow via its impact on banks' equity base.

The model has been simulated by means of agent-based techniques [17, 18]. The distribution of firms' size (in terms of capital stock) is characterized by persistent heterogeneity and it is well fitted by a power law, while the distribution of firms' growth rates can be approximated by a tent-shaped curve, i.e. a Laplace distribution. Both facts are consistent with the available evidence [11, 19]. Interestingly, the model replicates also the stylized facts for bankrupts discussed above. Simulations show that the distribution by age of firms that go out of business is to a good approximation exponential, while the right tail of the bad debt distribution scales as a power law with the scaling parameter α close to 1.

## 4. Conclusions

This paper presents evidence on firms' exit due to bankruptcy for a sample of European countries. We find that the right tail of the bad debt distribution scales as a power law with an exponent comprised between 0.7 and 1.1, and that the life-time of bankrupted firms is exponentially distributed. Our results are strikingly close to findings for Japan [14, 15], suggesting that there are universal features among the underlying microscopic mechanisms responsible for exits. We further discuss a recently proposed macroeconomic model with firms' financial fragility as a main ingredient, which is able to replicate empirical observation for bankruptcies and bad debt.